\documentclass[12pt]{iopart}

\begin{document}

\title[Chaplygin Gas Cosmology]{Chaplygin Gas Cosmology - Unification of 
Dark Matter and Dark Energy}
\author{Neven Bili\'c\footnote{Speaker}$^1$, Gary B Tupper$^2$ and Raoul D Viollier$^2$}
\address{$^1$Rudjer Bo\v{s}kovi\'{c} Institute, 10002 Zagreb, Croatia}
\address{$^2$Department of Physics, University of Cape Town,
Rondebosch 7701, South Africa}
\ead{bilic@thphys.irb.hr}


\begin{abstract}
The  models that unify dark matter and dark energy
based upon the Chaplygin gas  fail owing to the
suppression of structure formation by the adiabatic speed of sound. 
Including string theory effects, in particular the Kalb-Ramond field,
 we show how nonadiabatic perturbations 
allow a successful structure formation.
\end{abstract}



In contrast to the standard assumption that dark matter and dark energy
are distinct, there stands the hypothesis that both are different
manifestations of a single entity. The first definite model of this
type \cite{kame5,bil4,fab16} is based on the Chaplygin
gas, an exotic fluid with an equation of state
\begin{equation}
p = - A/\rho. 
\label{eq000}
\end{equation}
Subsequently, the generalization to
$p = - A/\rho^\alpha$, $0 \leq \alpha \leq 1$, 
was given
\cite{bent6} and the term `quartessence' coined \cite{mahl7}
to describe unified dark matter/energy models.

One of the most appealing aspects of the original Chaplygin gas model is that it is 
equivalent 
to the Dirac-Born-Infeld description of a Nambu-Goto membrane
\cite{bor,jac}.
However,
string theory D-branes possess three features that are absent in the simple 
Nambu-Goto membrane action: 
(i) they support an Abelian gauge field $A_{\mu}$ reflecting
open strings with their ends stuck on the brane, 
(ii) they couple to the dilaton,
(iii) they couple to the (pull-back of)
 Kalb-Ramond \cite{kalb20}
antisymmetric tensor field $B_{\mu \nu}$ which, like the gravitational 
field $g_{\mu \nu}$, belongs to the closed string sector.
Consider a $p$-dimensional D-brane with coordinates $x^{\mu}$,
$\mu=0,1...p$,
 moving in the p+1-dimensional bulk with coordinates $X^{a}$,
$a=0,1...p+1$. 
In the string frame the action is given by \cite{john} 
\begin{equation}
S_{\rm DBI}= - \sqrt{A} \:
\int d^{p+1}x\, e^{-\Phi}\sqrt{(-1)^p\det (g^{(\rm ind)} + {\cal B})}  \, , 
\label{eq001}
\end{equation}
where $g_{\mu\nu}^{(\rm ind)}$ is the induced metric
or the ``pull back" of the bulk space-time metric 
$G_{ab}$ to the brane,
\begin{equation}
g^{(\rm ind)}_{\mu\nu}=G_{ab}
\frac{\partial X^a}{\partial x^\mu}
\frac{\partial X^b}{\partial x^\nu} \, .
\label{eq002}
\end{equation}
The field ${\cal B}$ is an antisymmetric 
tensor field that combines
the Kalb-Ramond  and the electromagnetic fields
${\cal B}_{\mu \nu} = B_{\mu \nu} + 2\pi \alpha' F_{\mu \nu}$.
Let us 
choose the coordinates such that $X^\mu=x^\mu$
and let the $p+1$-th coordinate $X^{p+1}\equiv\theta$  be normal to the brane.
From now on, we set $p=3$ and consider a 3-brane universe in a 4+1 dimensional bulk.
Then
\begin{equation}
G_{\mu\nu}=g_{\mu\nu},
\;\;\;
\mu=0, ..., 3;
\;\;\;\;
G_{\mu 4}=0; \;\;\;\;
G_{44}=-1.
\label{eq005}
\end{equation}
After a few algebraic manipulations similar to \cite{gib},
the DBI action may be written as 
 \begin{eqnarray}
 &&
S_{\rm DBI} 
=\int d^4 x\sqrt{-\det g}\:
{\cal L}_{\rm DBI}\, ,
\label{eq106}
\\
&&
{\cal L}_{\rm DBI}= -\sqrt{A} e^{-\Phi}\, 
 \sqrt{(1-\theta^2)(1+ {\cal B}^2)
 -\theta {\cal B}^2\theta
 -({\cal B}\star{\cal B})^2 } ,
 \label{eq006}
\end{eqnarray}
where we have used the abbreviations:
\begin{eqnarray}
&&
\theta^2=g^{\mu\nu} \theta_{,\mu} \theta_{,\nu}, 
\;\;\;\;\;\;
{\cal B}\star{\cal B}=\frac{1}{8\sqrt{-\det g}}
\epsilon^{\mu\nu\rho\sigma}{\cal B}_{\mu\nu}{\cal B}_{\rho\sigma},
\nonumber
\\
&&
 {\cal B}^2=\frac{1}{2} {\cal B}_{\mu\nu} {\cal B}^{\mu\nu},
\;\;\;\;\;\;
\theta {\cal B}^2\theta=\theta_{,\mu}{{\cal B}^\mu}_\nu 
{\cal B}^{\nu\rho}\theta_{,\rho} \, .
\nonumber
\end{eqnarray}
Neglect for the moment the dilaton and the ${\cal B}$-field.
The DBI action (\ref{eq006}) then reduces to a scalar Born-Infeld  
action $S_{\rm BI} 
=\int d^4 x\sqrt{-\det g}\:
{\cal L}_{\rm BI}$ 
with the Lagrangian
\begin{equation}
{\cal L}_{\rm BI}=-\sqrt{A}\sqrt{1-\theta^2} .
\label{eq007}
\end{equation}
It may be easily shown \cite{bil4}
 that this Lagrangian yields the equation of state 
 (\ref{eq000}).

In a homogeneous model 
the conservation equation
${T^{\mu\nu}}_{;\nu}=0$
yields the density as
a function of the scale factor
$\rho (a) = \sqrt{A + B/a^{6}}$,
where $B$ is an integration constant.
The Chaplygin gas 
thus interpolates
between dust
($\rho \sim a^{-3}$) at large redshifts
and a cosmological constant
($\rho \sim \sqrt{A}$) today and hence yields a correct
homogeneous cosmology.

The inhomogeneous Chaplygin gas   based on
a Zel'dovich type approximation has been proposed \cite{bil4}, 
and the picture has emerged that on caustics, where the density is high,
the fluid behaves as cold dark matter, whereas
in voids, $w=p/\rho$ is driven to the lower bound $-1$ producing
acceleration as dark energy.
 Soon, however, it has been shown that the naive Chaplygin
gas model does not reproduce the  mass power spectrum \cite{sand8}
and the CMB \cite{cart9,ame}.

The physical reason is
that although  the adiabatic speed of sound,
defined by
$c_{\rm s}^2 = (\partial p/\partial \rho)_{S}$
is small until $a \sim 1$, the accumulated comoving acoustic horizon
\begin{equation}
d_{\rm s} = \int dt  c_{\rm s}/a \simeq H_{0}^{-1} a^{7/2}
\end{equation}
reaches Mpc scales by redshifts of twenty, frustrating the structure
formation even into a mildly nonlinear regime \cite{bil10}.
In the absence of caustic formation,
the density contrast described by a linearized evolution equation 
\begin{equation}
\ddot{\delta} + 2 H \dot{\delta} - \frac{3}{2}  H^{2} \delta 
-\frac{c_s^2}{a^{2}}  \Delta \delta=0  
\label{eq701}
\end{equation}
undergoes
damped oscillations that are in gross conflict with observations.
The root of the structure formation problem is the
last term on the left-hand side of (\ref{eq701}).
The solution in $k$-space
$\delta_k  = a^{-1/4}  J_{5/4} (d_{\rm s}k)$
shows the asymptotic behavior
\begin{equation}
\delta_k  \sim a \;\;\;\;\; {\rm for} \;\; d_{\rm s}k \ll 1 ;
\;\;\;\;\;\;\;
\delta_k \sim \frac{\cos d_{\rm s}k}{a^2} \;\;\;\;\; 
{\rm for} \;\; d_{\rm s}k \gg 1 .
\end{equation}
Hence, the perturbations undergo dumped oscillations at the scales
below $d_s$.

One simple way to save the Chaplygin gas is to suppose that nonadiabatic  
perturbations cause the
pressure perturbation $\delta p$ to vanish \cite{reis17}, and with it the acoustic horizon. 
 To achieve this, it is necessary to add new  degrees of freedom
\cite{bil6}
  which, to some extent, spoil 
 the simplicity of quartessence unification.

To illustrate this,
 consider the full
DBI action (\ref{eq106}) with (\ref{eq006}) which
contains extra degrees of freedom in terms of the dilaton $\Phi$
and the tensor field ${\cal B}$.
It is important to stress that all three fields, the dilaton, the scalar DBI field $\theta$,  and 
the Kalb-Ramond field
originate from an ultimate theory in the context of string/M theory and,
unlike in quintessence models, each field affects both dark matter and dark
energy. 
The full action 
\begin{equation}
S=\int d^4 x\sqrt{-\det g}\:\left( {\cal L}_{\rm b}
+{\cal L}_{\rm DBI} \right)
\end{equation}
contains, in addition to
the DBI action, the bulk terms 
 \begin{equation}
 {\cal L}_{\rm b}= \frac{1}{2\kappa^2} \,
 e^{-2\Phi} \left(-R - 4 g^{\mu\nu} \Phi_{,\mu} \Phi_{,\nu}+
\frac{1}{12} {\cal B}_{[\mu\nu,\sigma]} {\cal B}^{[\mu\nu,\sigma]}
\right),
\end{equation}
where $\kappa^2=8\pi G$.
It is convenient to write everything in Einstein's frame
using the transformation
$g_{\mu\nu}\rightarrow e^{2\Phi}g_{\mu\nu}$
and simultaneously rescaling 
$\theta_{,\mu}\rightarrow e^{\Phi}\theta_{,\mu}$ and 
 ${\cal B}_{\mu\nu}\rightarrow e^{2\Phi} {\cal B}_{\mu\nu}$.
In this way we obtain
\begin{eqnarray}
 &&
{\cal L}_{\rm DBI}= -\sqrt{A} e^{3\Phi}\, 
 \sqrt{(1-\theta^2)(1+ {\cal B}^2)
 -\theta {\cal B}^2\theta
 -({\cal B}\star{\cal B})^2 } ,
\\
&&
 {\cal L}_{\rm b}= \frac{1}{2\kappa^2} \,
 \left(-R + 2 g^{\mu\nu} \Phi_{,\mu} \Phi_{,\nu}+
\frac{1}{12} 
{\cal B}_{[\mu\nu,\sigma]} {\cal B}^{[\mu\nu,\sigma]}
+V(\Phi,{\cal B}) 
\right) ,
\\
&&
V(\Phi,{\cal B}) =
\frac{1}{3}\Phi_{[,\mu}{\cal B}_{\nu\sigma]}{\cal B}^{[\mu\nu,\sigma]} 
+\frac{1}{3}\Phi_{[,\mu}{\cal B}_{\nu\sigma]} 
\Phi^{[,\mu}{\cal B}^{\nu\sigma]} .
\end{eqnarray}
Applying the variational principle to ${\cal L}$, one
 may easily derive the energy momentum tensor
 \begin{equation}
T_{\mu\nu}=T^{\rm b}_{\mu\nu} +T^{\rm DBI}_{\mu\nu}
\end{equation}
and the equations of motion
for $\theta$, $\Phi$, and ${\cal B}_{\mu\nu}$.
Neither $T^{\rm b}_{\mu\nu} $ nor $ T^{\rm DBI}_{\mu\nu}$  
is in the form of a perfect fluid but we may still define
the corresponding $\rho$ and $p$ using the decomposition 
\begin{eqnarray}
& & T_{\mu \nu} = \rho u_\mu u_\nu - p h_{\mu\nu} + q_\mu u_\nu +
q_\nu u_\mu + \pi_{\mu \nu} \, ,            
\\
& & \rho = T_{\mu \nu} u^\mu u^\nu ,
\;\;\;\;
p =- \frac{1}{3}T_{\mu \nu}h^{\mu \nu} , 
 \\
 & &
q_\mu = T_{\nu \rho} u^\nu h^\rho_\mu \, ,
 \;\;\;\;
 \pi_{\mu \nu} = T_{\rho \sigma}h^\rho_\mu h^\sigma_\nu + ph_{\mu \nu}.
\label{eq523}
\end{eqnarray}
Here, $q_\mu$ and $\pi_{\mu\nu}$ are the  energy flux 
and the anisotropic pressure, respectively, and
$h_{\mu \nu} = g_{\mu \nu}-u_\mu u_\nu$
 is a projection
tensor. 

Next, let us see the implications for linear perturbation theory.
To study perturbations, it is convenient to work 
in the so-called temporal or synchronous gauge
 \begin{equation}
ds^2=dt^2-a^2(\delta_{ij}-f_{ij})dx^idx^j ,
\end{equation}
where $f_{ij}$ is treated as first-order departure from homogeneity.
In addition to that,  the density contrast 
$\delta = ( \rho - \bar{\rho} ) / \bar{\rho}$, and the field perturbations
 $\delta_{\theta} = \theta - \bar{\theta}$ and 
$\delta_{\Phi} = \Phi - \bar{\Phi}$ 
are first-order perturbations about the homogeneous and isotropic background.
The homogeneity and isotropy imply
$\bar{\theta}_{,i}=0$,
$\bar{\Phi}_{,i}=0$,
$\bar{\cal B}_{\mu\nu}=0$.
 As a consequence of this,
the spatial derivatives
$\Phi_{,i}=\delta{\Phi}_{,i}$ and
 $\theta_{,i}=\delta{\theta}_{,i}$,  
as well as the second order in ${\cal B}_{\mu\nu}$ 
 count as first-order
perturbations.

With these assumptions we may neglect higher-order
terms in the expression for the energy momentum
and we find simple expressions for the density and the pressure
associated with the DBI part $T^{\rm DBI}_{\mu\nu}$ of the energy momentum:
\begin{eqnarray}
 &&
\rho=T^{\rm DBI}_{\mu\nu}u^\mu u^\nu =\sqrt{A}e^{3\Phi}\sqrt{\frac{1+ 
{\cal B}^2}{1-\theta^2}}\, ,
\nonumber
\\
&&
p=-\frac{1}{3}T^{\rm DBI}_{\mu\nu}h^{\mu\nu}=-\frac{Ae^{6\Phi}}{\rho}
\left(1+\frac{1}{3}{\cal B}^2\right)\, .
\nonumber
\end{eqnarray}
Then, the pressure perturbation is given by
\begin{equation}
\delta  p  = - \bar{p}   \left(
\frac{\delta \rho}{\bar{\rho}}
 + 6 \delta \Phi -\frac{1}{3}{\cal B}^2\right)\,
\label{eq501}
\end{equation}
and the nonadiabatic cancellation scenario  
 is realized by 
 \begin{equation}
\frac{\delta\rho}{\bar{\rho}}
 + 6 \delta \Phi -\frac{1}{3}{\cal B}^2=0
\end{equation} 
 as an initial condition outside the causal horizon
$d_{c} = \int dt / a \simeq  H_{0}^{-1}  a^{1/2}$. 
However, it is essential that once the perturbations enter the 
causal horizon $d_c$,
 at least one of the two components,
 $\delta\Phi$ or 
${\cal B}^2$, grows in the same way as the density contrast.

In the following  we  demonstrate that whereas the dilaton does not
  yield  the desired cancellation of 
 nonadiabatic perturbations, the Kalb-Ramond field might provide such a
mechanism.

\subsection*{The Dilaton}
Retaining only the dominant terms, the dilaton perturbation 
$\delta \Phi$ satisfies \cite{bil18}
\begin{equation}
\delta \ddot{\Phi} + 3 H \delta \dot{\Phi} -
\frac{1}{a^{2}}  \Delta \delta \Phi \simeq 0 \,   ,
\label{eq301}
\end{equation}
with the solution in $k$-space
\begin{equation}
\delta \Phi_k = a^{-3/4}  J_{3/2} (kd_{\rm c}) .
\label{eq302}
\end{equation}
Then, once the perturbations enter the causal horizon $d_{c}$ 
(but are still outside the
acoustic horizon $d_{s}$), $\delta \Phi$ undergoes rapid damped
oscillations, so that nonadiabatic perturbation associated with 
$\Phi$ is destroyed. 
This means that the nonadiabatic
 perturbations are not automatically preserved except at long,  i.e.,
superhorizon, wavelengths where the naive Chaplygin gas has no problem anyway.

\subsection*{The Kalb-Ramond Field} 
\label{kalb}

To estimate the effect of the ${\cal B}$-field, we 
addopt the temporal gauge ansatz \cite{bil6,chun21}
\begin{equation}
{\cal B}_{0i} = 0,
\;\;\;\;\;
{\cal B}_{i j} = \epsilon_{i j k}  B^{k} .
\end{equation}
Then the ${\cal B}$-field term in (\ref{eq501}) 
is
\begin{equation}
{\cal B}^2=\frac{B^iB^i}{a^4}.
\end{equation}
Retaining the dominant terms, the field equation 
for $B^{i}$ takes the form
\begin{equation}
\frac{d}{dt}\left(\frac{{\dot B}^i}{a}\right)-
\frac{B^j_{,ji}}{a^3}
+
\frac{2 \kappa^2 A}{\bar{\rho}} \: \frac{B^i}{a} = 0\,  .
\label{eq014}
\end{equation}
If we make the decomposition
 $B^i=B^i_{\bot}+B^i_{\|}$ 
 with the transverse part
satisfying 
$\partial_i B^i_{\bot}=0$,
the key point becomes evident: whereas the longitudinal part
  $B^{i}_{\|}$ suffers the
same problem as $\delta\Phi$ in  (\ref{eq301}),
the transverse part does not experience spatial gradients.

As $H^{2} = \left( \dot{a} / a \right)^{2} = \kappa^2 \bar{\rho} / 3$,
the last term in (\ref{eq014}) is of the order
$H^{2} A/ \bar{\rho}^{2}$, so being
negligible compared with the first term which is ${\cal{O}} (H^{2})$,
until $a \sim 1$. Then, equation (\ref{eq014}) simplifies to
\begin{equation}
\frac{d}{dt}\left(\frac{{\dot B}^i}{a}\right)-
\frac{B^j_{,ji}}{a^3}=0.
\label{eq401}
\end{equation}
First, we show that 
once the perturbations enter the causal horizon,
the longitudinal part of 
$B^iB^i/a^4$ in 
 (\ref{eq501}) undergoes damped oscillations,
similar to those experienced by $\delta \Phi$.
The longitudinal component of (\ref{eq401}) in 
$k$-space reads
\begin{equation}
\frac{d}{dt}\left(\frac{{\dot B}_{\|}(k)}{a}\right)+
\frac{k^2}{a^3}B_{\|}(k)=0 ,
\label{eq406}
\end{equation}
with the solution
\begin{equation}
B_{\|}(k)=a^{5/4} J_{5/2}(kd_{\rm c})
\end{equation}
Hence 
$B_{\|}$ oscillates inside
 $d_{\rm c}=2\Omega^{-1/2}
H_0^{-1}a^{1/2}$
 with the amplitude 
increasing linearly with $a$.
As a consequence,
the longitudinal term $B_{\|}^2/a^4$  that enters 
the right-hand side of (\ref{eq501}) oscillates with an amplitude decreasing  
as $a^{-2}$.
 
 Next we show that  
the transverse term $B^i_{\bot}B^i_{\bot}/a^4$ 
grows linearly.
In the transverse components of (\ref{eq401}) 
the last term is absent, 
 so the transverse solution reads
\begin{equation}
B_{\bot}^{i} ( a, \vec{x} ) \simeq c^{i} ( \vec{x} ) 
\int_{0}^{a} \frac{da}{H} = \frac{2}{5} \: \frac{c^{i} 
(\vec{x})}{H_{0} \Omega^{1/2} } \: a^{5/2} ,
\end{equation}
where $\Omega$ is the equivalent matter fraction at high redshift
and $c^{i} ( \vec{x} )$ are arbitrary functions of $\vec{x}$.
Thus,
\begin{equation}
\frac{B_{\bot}^{i} B_{\bot}^{i}}{a^4} = \frac{4}{25} \:
\frac{c^{i} c^{i}}{H_{0}^{2} \Omega} \: a
\label{eq402}
\end{equation}
grows linearly with the scale factor. 
Owing to this equation
  we can arrange
nonadiabatic
perturbations  such that
\begin{equation}
B_{\bot}^{i} B_{\bot}^{i} / a^{4} - 3 \delta = 0
\end{equation}
and hence $\delta p=0$
with the assurance that this will hold independent of scale
until $a \sim 1$. 
Here we denote the density contrast by
$\delta=\delta\rho/\bar{\rho}$
for overdensities only. With vanishing $c_s^2$, the spatial gradient term
in (\ref{eq701}) is absent, and 
the density contrast satisfies
\begin{equation}
\ddot{\delta} + 2 H \dot{\delta} - \frac{3}{2}  H^{2} \delta = 0 ;
\hspace{1cm} \delta > 0
\end{equation}
with the growing mode solution
$ \delta \propto a$ .
This  is our main result: the growing mode overdensities here do
not display the damped oscillations of the simple Chaplygin gas below
$d_{\rm s}$, but grow as dust. We remark that it matters little that this
applies only for $\delta > 0$ since the Zel'dovich approximation
implies that 92\% ends up in overdense regions.

Clearly, there is an open question as to what inflation model can produce the
initial conditions in the Kalb-Ramond field and the brane embedding to allow
subsequent structure formation. We believe this issue is likely to be
closely related
to another: namely, how does the Chaplygin-Kalb-Ramond model fit into the
braneworld picture? 
The estimate presented here can be taken as a starting point
for investigating questions beyond linear theory, 
in the complete general-relativistic perturbation analysis
including the electric-type field ${\cal B}_{0i}$.
 In particular, one might
hope that the acoustic horizon does resurrect at very small scales to provide
the constant density cores seen in galaxies dominated by dark matter.

Ultimately, the model must be confronted with large-scale structure
 and the CMB.
 In particular, the Kalb Ramond field may leave a mark
 on the power spectrum and the CMB spectrum. To investigate 
 that, one would need to do the full perturbation analysis, 
which would go beyond the scope of the present paper.
 For the simple Chaplygin gas where the standard model feels the 
metric $g_{\mu \nu}$,
it has been shown \cite{bil22} that a good fit to the data
is obtained if a vanishing sound speed
is imposed on the Zel'dovich fraction.

\vspace{0.2in}

The work of NB   was supported by
 the Ministry of Science and Technology of the
 Republic of Croatia under Contract
 No. 0098002.

\end{document}